# Assisted morbidity coding: the SISCO.web use case for identifying the main diagnosis in Hospital Discharge Records

Elena Cardillo[*], Lucilla Frattura[**]

**Abstract:** Coding morbidity data using international standard diagnostic classifications is increasingly important and still challenging. Clinical coders and physicians assign codes to patient episodes based on their interpretation of case notes or electronic patient records. Therefore, accurate coding relies on the legibility of case notes and the coders' understanding of medical terminology. During the last ten years, many studies have shown poor reproducibility of clinical coding, even recently, with the application of Artificial Intelligence-based models. Given this context, the paper aims to present the SISCO.web approach designed to support physicians in filling in Hospital Discharge Records with proper diagnoses and procedures codes using the International Classification of Diseases ($9^{th}$ and $10^{th}$), and, above all, in identifying the main pathological condition. The web service leverages NLP algorithms, specific coding rules, as well as ad hoc decision trees to identify the main condition, showing promising results in providing accurate ICD coding suggestions.

## 1. Introduction

The proper use of standard classifications, such as the International Classification of Diseases (ICD) and coding of morbidity data has always been fundamental for all general epidemiological and many health-management purposes (WHO, 2016). One example is the use of the information flow of the Hospital Discharge Records (SDO) collected in national databases for monitoring hospitalization episodes provided in public and private hospitals and thus the provision of hospital assistance. This has become an indispensable tool for both administrative analyses (i.e., for accurate billing) and clinical elaborations (e.g., health quality assessment), which can bring to the planning of new measures to support healthcare and welfare activities or to more strictly clinical-epidemiological and outcome analyses.

In this frame, although approaches to coding vary across institutions, clinical coding specialists frequently perform coding retrospectively. The assignment of codes to each patient episode of care during hospitalization is determined by different factors, among others by the coder's interpretation of the available case notes or the completeness of the electronic health records. As a result, accurate coding is dependent on both the intelligibility of the case notes and the coders' knowledge of medical terminology (Sundararajan *et al*. 2015).

Several studies have indicated poor reproducibility of clinical coding (Tatham A., 2008) and poor accuracy which seems not dependent on the version of the standard coding system used, which in the case of SDO is ICD (Quan *et al.* 2014).

In recent years, even if the application of artificial intelligence (AI) has begun to attract and, in some cases, assist clinicians in the practice of medical coding, the performances achieved by AI models do not meet expectations. Many studies have proven this, especially concerning inadequate levels of data coding accuracy (less than 50%) and high computational costs (Falis *et al.* 2024; Soroush *et al.* 2024). This means that more reliable and trustworthy systems are required to support physicians or coders in speeding up the coding process while retaining the necessary precision.

---

[*] Institute of Informatics and Telematics, National Research Council, Rende, Italy. elena.cardillo@iit.cnr.it
[**] Azienda Sanitaria Universitaria Giuliano Isontina (ASUGI), Udine, Italy. lucilla.frattura@asugi.sanita.fvg.it

Given this context, the paper aims to describe the results of the "SISCO.web" project[1], whose scope was the design and implement a Coding Support System (CSS), in the form of a web service, to improve accuracy in coding health conditions in Italian Hospital Discharge Records (SDO). The main objective of the service is to support Italian physicians (coders) in morbidity coding, and more specifically in the coding of diagnoses and procedures/interventions using ICD-9th revision, Clinical Modifications (ICD-9-CM), mandatory in Italy, and, more notably, in identifying the "main condition" to be filled in SDOs.

The paper is structured as follows: Section 2, provides background information on using and coding SDO, and describes the applied methodology. Section 3 showcases the results and includes a preliminary evaluation. Section 4 presents some related works, and finally, Section 5 offers conclusions and future directions.

# 1. 2. Materials and Methods

## *2.1.* Hospital Discharge Records

The Hospital Discharge Record Database was established, in Italy, with the decree of the Ministry of Health on 28 December 1991. It serves as a tool for collecting information about each patient discharged from public and private hospitalization institutions across the country. The information gathered in each SDO includes, beyond the patient's characteristics (e.g., age, sex, etc.), the peculiarities of the hospitalization (e.g., institution and discharge discipline, method of discharge, etc.) and, above all, clinical features (e.g., the main diagnosis, concomitant diagnoses, diagnostic or therapeutic procedures, and interventions), excluding information relating to drugs administered during hospitalization[2].

Subsequently, other decrees introduced new regulations for the information flow transmission to the Ministry of Health, expanded the information content of the SDO, and adopted the international classification ICD-9-CM version 1997 (Ministerial Decree October 27, 2000, n. 380) for the coding of diagnoses and diagnostic and therapeutic procedures, then updating this regulation with the adoption of the 2007 Italian version and introducing the adoption of the Diagnosis Related Group classification (DRG), version 24 for hospital admissions (Ministerial Decree December 12, 2008).

In 2011, the "It.DRG Project", coordinated by the Ministry of Health was launched to develop a new classification and assessment method for inpatient care, specific and representative to the Italian context (Sforza *et al.* 2021). The objective of this project was: i) the development and testing of an updated version of the ICD-10 classification (International Classification of Diseases and Health Related Problems-10th Revision) that incorporates WHO-approved updates and makes minor changes, finalizing the so-called Italian modification of ICD-10 (ICD-10-IM); the development and testing of the Italian classification of Interventions and Procedures (CIPI), a version of the section on procedures and interventions of ICD-9-CM modified and supplemented, to adapt it to specific Italian needs and to provide for integration with codes that allow for the detection of information on (i) Procedures/treatments provided (also) in ambulatory care; (ii) Medical-surgical devices; (iii) High-cost drugs; and iii) finally, a new version of the DRG system (Nonis *et al.* 2018).

---

[1] The "SISCO.web" project, funded by the Friuli Venezia Giulia (FVG) Region and coordinated by the Italian Collaborating center of the World Health Organization Family of International Classifications (WHO-FIC) in Udine through the Azienda Sanitaria Bassa Friulana Isontina n. 2 (incorporated now into the "Azienda sanitaria universitaria Giuliano Isontina" - ASUGI) was executed from 2017 to 2021 and led to the development of a prototype (SISCO.web service) which can assist clinicians in coding SDO data to using ICD-9-CM, but it is also set up to support ICD-10 coding.

[2] Hospital Discharge Records database (HDR/SDO), see for details: https://www.healthinformationportal.eu/health-information-sources/hospital-discharge-database-2

Despite the significant outcomes of the "It.DRG project" for innovating and improving SDO data management, there is a need to create a roadmap for implementing the new classifications, especially ICD-10, in a more simplified manner. This involves using crosswalking tables to ICD-10-IM and confirming the planned current version of DRG classification. The attention in this paper is paid primarily to a tool for coding diagnoses and intervention using ICD-9-CM, with the understanding that the mentioned crosswalking tables for coding diagnoses in ICD-10-IM can be easily implemented in the tool's architecture.

*2.1.1. The International Classification of Disease*

The International Classification of Disease is the most known and widely used standardized WHO classification system, which was originally intended to facilitate the statistical analysis of health data (Moriyama *et al.* 2011). Each successive revision to the ICD, typically spanning 10-20 years, has sought to address new use cases while adapting to advances in medicine and healthcare and has continued to grow in number of total codes (Williamson *et al.* 2024). The tenth version, ICD-10 has approximately 14.000 codes for health conditions, signs, symptoms, and reasons to encounter health services. This revision has then been renewed with the implementation of the eleventh revision of the classification, ICD-11[3], developed thanks to an unprecedented collaboration between WHO working groups, knowledge engineers and informaticians from Stanford University (USA), and professionals all over the world to become a global standard for health data clinical documentation and statistical aggregation. It presents a new coding structure compared to previous revisions and is fully digital for the first time. The basic component is an underlying ontology database containing all ICD entities (over 55,000 unique entities)[4]. The new structure, its digital nature, and the tools provided to support the use of the classification enhanced its application flexibility. Moreover, it is interoperable with health information systems and other coding systems.

As mentioned above, in Italy, ICD-9-CM is used for morbidity coding, containing over 15,000 diagnosis codes. Its use is also recommended in primary care prescription documents and for diagnoses and problems encoding in the Italian Patient Summary[5] each entity within the ICD-9-CM is encoded by a unique identification string consisting of three to five digits and an optional single letter prefix corresponding to a supplementary category. Practical applications of the ICD in healthcare have expanded and now have come to include the indexing of health record data in hospitals, the coding of medical billing claims (Moriyama *et al.* 2011), and the assessment of quality of patient care (O'Malley *et al.* 2005).

*2.1.2. The coding of the main condition*

A coded health data record can have a varying number of diagnostic codes. Some authors, considering that there is no uniform definition of "main condition", noted that one of these diagnoses must be coded as the main condition, known also as "main diagnosis", "primary diagnosis", "principal diagnosis or "discharge diagnosis" (Sukanya, 2017).

Two definitions have been used for the main condition in ICD-coded health data: a "resource use" definition and a "reason for admission" definition. In Italy, the first definition is implemented, as said above, in detecting and coding the discharge diagnosis using ICD-9-CM, 2007 version (Italian Ministry of Labor, Health and Social Affairs, 2008). In the Italian SDO, it is necessary to code the main diagnosis, and several other diagnoses related to the hospital episode of care. The mentioned national database on SDO contains more than 290 million records (7.957.647 only in 2023). Annual reports are available for download from the website of the Italian Ministry of Health (Italian Ministry of Health, 2022). Coding of these records is made directly by clinicians and health professionals, with some levels of accuracy monitoring at the hospital and regional level before the data are sent to the Ministry of Health periodically. This richness of data must face with its accuracy. Several Italian studies are available showing low accuracy in coding. Hospital discharge data were found to be specific but insensitive in many fields. For example, the reporting of acute ischemic stroke and thrombolysis provides misleading indications about both the quantity and quality of acute ischemic stroke hospital care in many studies (see Rinaldi *et al.* 2003; Spolaore *et al.* 2005). Other studies show that Hospital discharge records appear to poorly reflect the incidence of amyotrophic lateral sclerosis and can be used only after clinical verification of the diagnosis (Chiò *et al.* 2002). Moreover, looking at (Amodio *et al.* 2014), the diagnosis of influenza seems to be overcoded. Nevertheless, based on the retrieved evidence, administrative databases can be employed to identify primary breast cancer. The best algorithm suggested is ICD-9 or ICD-10 codes located in the primary position (Abraha *et al.* 2018). At an international level, many studies confirmed that physicians do not code the disease in SDOs according to the main diagnosis principles (Wang *et al.* 2021). It is observed that in many cases, the main diagnosis is mistaken for an outpatient diagnosis, making it more difficult to identify when multiple diseases occur simultaneously or in cases of complications. These studies reveal that physicians still require support to collect, classify, analyze, and use medical record information according to disease classification criteria.

---

[3] International Classification of Diseases, Eleventh Revision (ICD-11), World Health Organization (WHO) 2019/2021: <https://icd.who.int/browse11>.

[4] These entities include diseases, injuries, external causes, signs and symptoms, substances, drugs, anatomy, etc., pointing to about 17.000 categories, for over 120,000 clinical terms covered, allowing the description of health conditions at any level of detail by combining codes.

[5] Italian Permanent working table for Digital health in Regions and Autonomous Provinces. "Specifiche tecniche per la creazione del "profilo sanitario sintetico" secondo lo standard HL7-CDA rel. 2". Department for the Digitization of Public Administration and Technological Innovation, 2010.

## 2.2. The SISCO.web approach

The main scope of the SISCO.web service, as mentioned above, is to support the coding of SDOs, guiding the physicians to identify and code the main condition, allowing the most appropriate ICD-9-CM codes, and in the future ICD-10 codes. This means that its function is to guide the user before the compilation of the SDOs, to choose and assign appropriate ICD codes to the diagnostic formulations available in medical record documentation collected during patient hospitalization, and, further, to identify among different diagnoses, the main one (Cardillo *et al.* 2019). Peculiarities of this support system are:

- A knowledge base containing clinical concepts, related terms, and mappings to ICD-9-CM for managing the transition from the usual scientific language to the coding language. This means, the integration of such resources with the ICD-9-CM systematic index, the ICD-9-CM alphabetical index, and other additional terms (synonyms, acronyms, linguistic variants, common medical terms, etc.).
- Standardized coding rules (e.g., "diagnostic and procedure codes are to be used at their highest level of specificity"; "three-digit codes are to be assigned only if there are no four-digit codes within that code category"; etc.).
- A rule engine for managing these rules, represented by the Business Rules Management System (BRMS) "Drools".

As shown in Fig. 1, the SISCO.web architecture includes three main layers:

1. Presentation layer: handling the interactions that users have with the software. Here the web component, has a multi-tier architecture, deployed on a Tomcat web server, offering two web interfaces (WUIs) to support the compilation of SDOs. The WUIs make JSON calls to the Web Services of the underlying levels, which access the data resources built by the batch component. The two WUIs allow for two specific tasks: i) the text encoding WUI (TEM module), which serves as a coding tool, since it allows for searching clinical terms (diagnoses and procedures) and suggests the most appropriate ICD-9-CM codes based on search algorithms and related terms derived from the knowledge base; ii) the identification of the main diagnosis WUI (IMDM module), based on a rule engine that implements a specific decision tree for choosing and coding the main condition among the multiple diagnoses selected in the previous step. These two modules will be described in detail in the following paragraphs.

2. Application layer: handling the main code definitions and the most basic functions of the developed application. In SISCO.web this layer includes five main functions which will be detailed later (e.g., search, autocomplete, code Details, use of related Terms for improving search, coding rules application through the Drools engine).

3. Data layer: which is mainly devoted to data storage. In fact, it houses not only data but indexes and tables. Here the batch component is aimed to build the data resources, i.e., the SISCO.web knowledge base, which is stored on the Apache Lucene Index.

The Apache Lucene Index[6], was chosen because it is a valid open-source tool for retrieving data and information. It provides straightforward Java APIs for creating text indexes and full-text search with options such as proximity search, fuzzy search, and score-based sorting, weighted filter search (meaning that each filter can be allocated a variable weight for scoring purposes in the search results).

To implement the RESTful layer of web services within the system architecture, we chose Jersey[7]. This open-source framework is based on the JAX-RS API and uses annotation-based programming, which simplifies the creation of RESTful web services. Jersey also facilitates the representation of data in standard formats such as JSON, XML, and HTML.

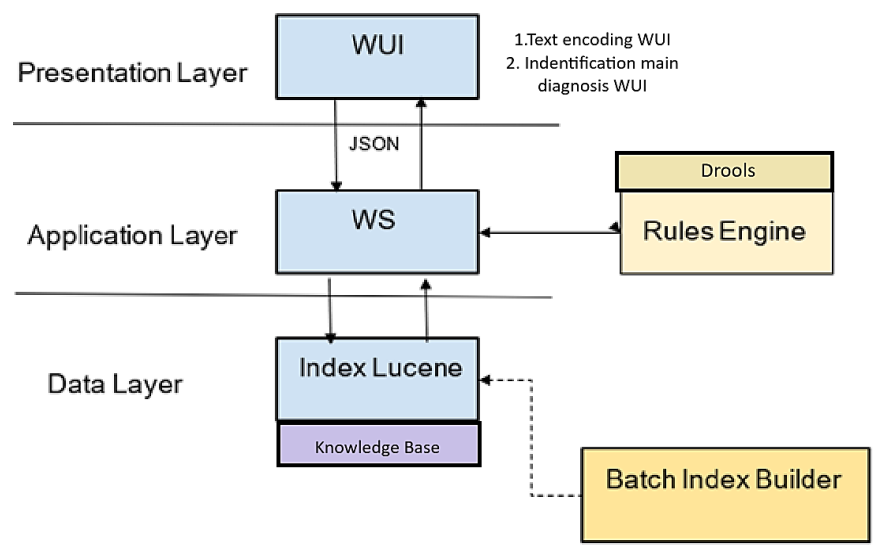

**Figure 1.** SISCO.web architecture

The main process to reach the supported coding of morbidities and procedures and the identification of the main condition is shown in Fig. 2 and can be briefly described as follows:

1. Using the first module, i.e., TEM, the user starts searching for a diagnosis (one at a time) using the ones reported in the discharge letter (LDO) of the patient, to look for its ICD-9-CM code.

---

[6] Apache Lucene is available for download at: https://lucene.apache.org/
[7] Eclipse Jersey is available for download at: https://eclipse-ee4j.github.io/jersey/

2. The system applies classic NLP algorithms such as Tokenization, text similarity algorithms to assign the most appropriate code to the diagnosis plus Decision Trees, and Symbolic NLP algorithms, i.e., rule-based and knowledge-based algorithms, relying on predefined linguistic rules and knowledge representations. For this reason, dictionaries, grammars, and ontologies are used to process language.
3. Every time the user searches for a diagnosis and selects one of the results suggested by the system, a list of coded diagnoses is generated to allow the user to identify, among these diagnoses the main condition.
4. The same procedure is used to search for procedures and interventions if reported in the discharge letters, and a second list of coded procedures/interventions will be generated by the system to be used as well by the IMDM module.
5. These two lists of codes represent the input data for the decision tree algorithm, which, as described in Subsection 2.2.2., will guide the user to identify the main pathological condition based on the defined coding rules.

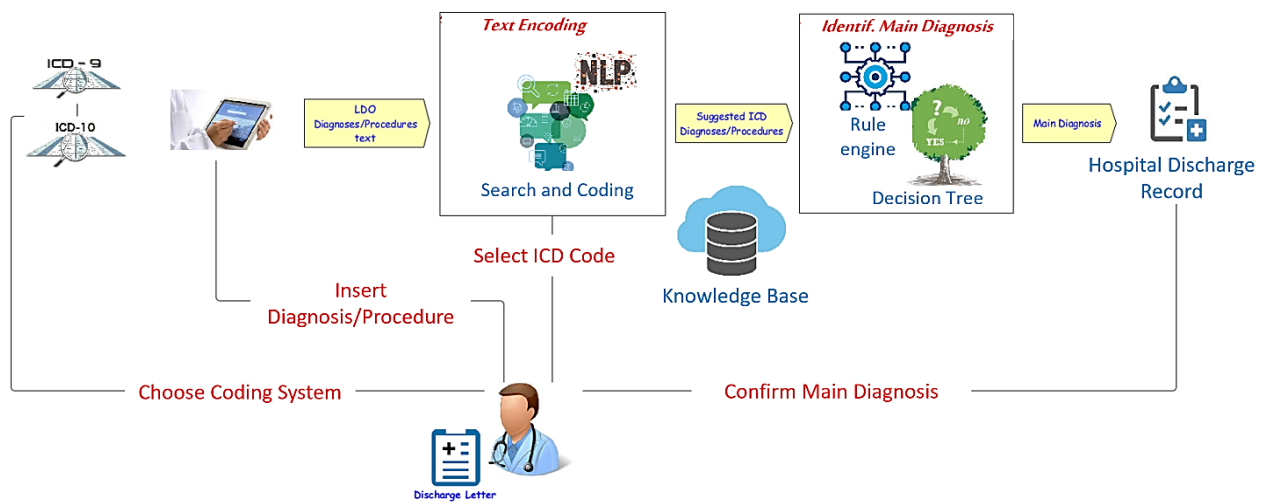

**Figure 2.** The SISCO.web main process

To better understand the above-mentioned process, the next subsections will give details on the knowledge base, the algorithms, the decision tree and the coding rules used in the two modules to suggest the most appropriate ICD-9-CM codes.

### 2.2.1. The SISCO.web Knowledge Base

The knowledge base (KB) built for the project and used in the TEM integrates a series of terminological resources related to diagnoses and interventions/procedures in EHRs. The main data sources, as shown in Fig. 3, are represented by the Italian versions of:

- ICD-9-CM (v. 2007), Systematic index of diagnoses and procedures, considering the codes at the maximum level of specification.
- ICD-9-CM (v. 2007), Alphabetic index of diagnoses, and Alphabetic index of procedures.

For this project, an ontological version of ICD-9-CM has been created starting from the available ministerial tables of the classification, bringing to the development of the ICD-9-CM Ontology in Owl.

The lists of terms present in ICD-9-CM, in some cases inappropriate or outdated jargon, were supplemented with terms taken from other sources such as:

- Ad hoc created glossaries of diagnoses derived from physicians' scientific language, developed during a previous project (see Cardillo *et al.* 2018).
- A glossary of diagnoses coded in ICD-9-CM extracted from the FVG Emergency Department (ED) EHRs database.
- Rare Diseases terms (Prime Minister's Decree 2017).
- Italian MeSH diagnoses and procedures terms[8].

All the terms derived from these sources were in most cases already mapped to the corresponding ICD-9-CM codes and were qualified as exact or approximate mapping.

Regarding the resource extracted from FVG Emergency Department "SEI Database", in the beginning, a list of 425 common pathological conditions in the ED was proposed by the ED FVG regional working group. On this list, a further analysis was performed to verify the use of technical/scientific terms and the correctness of the ICD-9-CM coding associated to these pathological conditions, bringing in the end to a glossary of 696 diagnoses (2,530 words) which enriched the SISCO.web KB.

**Table 1.** Knowledge Base SISCO.web (Cardillo *et. al*, 2019)

| Resources | Version | N. of Terms |
|---|---|---|
| ICD-9-CM systematic index | IT- 2007 | 16,294 |
| ICD-9-CM alphabetical index | IT- 2007 | 289,834 |
| Physicians' Glossary of diagnoses | v. 2017 | 1,421 |
| Rare Diseases terms | v. 2017 | 683 |
| Emergency physicians' diagnoses and pathological conditions (SEI database) | v. 2018 | 696 |
| MeSH synonyms for diagnoses and procedures | v. 2017 | 641 |
| Neoplasms related terms | v. 2017 | 13,290 |
| **Total** |  | **322,859** |

As observable, the total number of terms in the KB, considering the whole Italian ICD-9-CM resource and the above-mentioned additional resources, is about 323,000. It's important to note that the entire SISCO.web KB, particularly the data extracted from the SEI dataset, is not publicly accessible.

Regarding the ICD-9-CM Ontology, as said above, we created a processable version of the Ministerial file published online, since the original .xls file missed important details about each ICD-9-CM code. This information includes descriptions, inclusion and exclusion criteria, and notes, which are crucial for giving coding support based on ICD. To solve this problem, we developed a script that builds a lightweight ontology in Owl which can also be used to search for inconsistencies in the ICD-9-CM hierarchy or in the attribute's association. The ontology classes are based on the structure of the ICD-9-CM systematic index. At the top level, there are two main classes representing the ICD-9-CM main sections: *Diseases and Injuries*, and *Procedures and Interventions*. Within the *Diseases and Injuries* section, there are 17 classes that correspond to the ICD-9-CM "chapters" in this category, along with two additional classes for supplementary classifications: one for external causes of injury and poisoning and another for factors influencing health status and contact with health services.

Each chapter has its own class hierarchy, following the index structure that includes *blocks*, *categories*, *subcategories*, and *subclassifications*. To help with navigation, we labelled chapter classes with chapter numbers (e.g., Chapter I, Chapter II) and use E and V for the additional classes.

---

[8] Medical Subject Headings Thesaurus. Available at: <https://old.iss.it/site/Mesh/>.

Similarly, in the *Procedures and Interventions* section, each category is organized under ranges such as the *Nervous System Intervention*, which covers codes 01-05. Each class/subclass in the ontology connects to the relevant data type annotations and, when needed, to Object Properties (i.e., relationships between classes) and axioms. Access to the ICD-9-CM Ontology is currently restricted. However, we are planning to make it available on public repositories or GitHub shortly.

*2.2.1. The Text encoding module*

The first module is designed for searching the appropriate code for one or more diagnoses and procedures/interventions mentioned in the patient's discharge letter. The user enters a diagnosis in the search box using free text, which can be a single word or a multi-word term (T1). As the user starts typing, the system provides suggestions for autocompletion based on the knowledge base (KB), drawing from systematic or alphabetical indexes, MeSH synonyms, glossaries of general practitioners or emergency physicians, rare diseases, etc. These suggestions are the ones that have the entered text as their prefixes. Subsequently, the system conducts a syntactic search on the description of each attribute associated with ICD-9-CM classes in all types of resources in the KB. Different weights are assigned to each attribute based on its source and position. The search yields a list of ICD classes (diagnoses/procedures) that meet the search criteria, i.e., one or more attributes containing T1. The results are displayed in descending order based on their score.

To enhance the search function for the coder, the system permits filtering of the results in the list. This is achieved by incorporating the terms used in the query with related terms suggested by the system. These suggestions are based on their co-occurrence with the searched term within the ICD descriptors. To be more specific, the descriptions of the resulting ICD classes are tokenized to extract the most significant words (stop words are not considered). Moreover, to facilitate the tokenization and subsequent counting of term occurrences, the following ICD attributes are to be considered:

- The main description of the ICD class, along with any supplementary descriptions and inclusion terms in the systematic index.
- The description of the entry terms in the alphabetical index.

The system counts the number of times each token/term appears in the list of ICD classes resulting from the search. It then arranges the terms in descending order based on the number of occurrences and presents them to the user as related terms in a separate box. The user can choose one of the related terms or continue entering other free text in the search box. The system provides suggestions for autocompletion as the user enters more terms (T1, T2, etc.). The result list of ICD codes (diagnoses or procedures, depending on the user's initial selection) is updated to consider the search criteria, ensuring that one or more attributes contain all the input terms (T1, T2, etc.), and co-occurrences, making the search more precise. A similar approach is used in the ICD-11 Coding Tool[9], which, unlike SISCO.web, allows also to use ICD chapters and ranges as research filters. From here on the algorithm performs the same steps, until the user selects a specific diagnosis/procedure among the ICD classes displayed in the search results which is always a leaf code. Once the diagnosis/procedure is selected, the system adds it to the list of candidate diagnoses/procedures used by the decision tree algorithm for identifying the main condition.

Is worth mentioning that the search and coding algorithm for procedures follows the same steps as that for diagnoses, but the Knowledge base which supports the process is smaller. In fact, in the case of procedures, the NLP algorithm examines only terminological resources related to interventions and procedures, therefore fewer terms are indexed. Specifically, the search is conducted almost entirely on the classes contained in the systematic index of ICD-9-CM section procedures, as well as on the procedure terms present in the ICD-9-CM alphabetical index, and the external resource MeSH (for the terms related to procedures). Consequently, the search algorithm is, in some way, simpler than that utilized for diagnosis searches.

---

[9] ICD-11 Coding tool is used to find the correct ICD-11 code for a specific diagnosis and it is connected to the ICD-11 browser to allow user to see further details for a searched diagnosis. It is available at the link: < https://icd.who.int/ct/icd11_mms/en/release>.

*2.2.2. The Identification of the main diagnosis module*

To support physicians in the identification of the main condition, a decision tree was created to adhere to the WHO guidelines for morbidity coding in ICD-10 (Zavaroni *et al.* 2018). This includes following, on one hand, the WHO ICD-10 rules and guidelines for morbidity coding (WHO, 2016)[10], which are up-to-date compared to ICD-9-CM 2007 rules, and on the other hand the WHO definition of the main condition, i.e., « the condition, diagnosed at the end of the episode of health care, primarily responsible for the patient's need for treatment or investigation». (WHO, 2016, p.147)

Furthermore, interventions and procedures were also considered in the decision-making process. To manage the extensive array of ICD codes (about 5,000), they were grouped into three sets:

1) "relevant surgery": encompassing interventions or procedures typically requiring an operating room, or those with resource consumption comparable to operating room costs;
2) "selected non-relevant surgical interventions": encompassing interventions or procedures, other than relevant surgery, that require significant resources, mostly higher than a non-surgical treatment of a condition;
3) "residual non-relevant surgical interventions": encompassing interventions or procedures that necessitate fewer resources than non-surgical treatments.

Conditions were categorized into "conditions" (including diseases and clinical manifestations or normal physiological changes) and "pathological conditions" (abnormal anatomy or functioning constituting diseases).

The decision tree hierarchy includes: i) specific hospital settings which are highly specialized by age and changes of particular conditions, such as "neonatology" and "pregnancy, delivery, and puerperium", foreseen specific or partially specific paths; ii) paths for the other hospital settings, according to the general rules and, iii) the interventions/procedures set. Notably, the third group of interventions/procedures mentioned above is excluded as a viable option for identifying the main condition.

In Summary, the coding of certain health conditions is driven by the condition itself (pregnancy and related conditions, neonatal health), whereas for others, resource consumption due to procedures is the primary determinant. Thus, when a relevant intervention/surgery is identified, it influences the choice of the targeted condition. The decision tree rules are integrated into the rule engine module of the SISCO.web service.

The algorithm which determines the main condition, uses a Drools-based rule engine. Drools is an open-source Business rule management system (BRMS), released under the Apache License 2.0., that can easily be embedded in any Java application, which include an inference engine based on forward and backward chaining (Proctor 2012). The primary function of the Drools rule engine is to match incoming data, (i.e., facts), to the conditions outlined in the rules. It then determines whether and how to execute these rules. Key components in Drools are the following: *rules*; *facts* that are matched against the conditions of the rules to execute the applicable ones; a *production memory* (i.e., where the rules are kept); a *working memory* (i.e., location for the facts)[11].

In our implementation, the system consists of four components, developed in Java, and utilizes the RabbitMQ message broker (see Fig. 4). The primary component is the SISCO Drools Engine, serving as a wrapper for the Drools engine. It takes input data that triggers the execution of one or more rules down to a node, corresponding to a decision (leaf node), or the generation of a request for other parameters. The modules exchange messages in JSON format. On the web server side, the SISCO Rules Web Service component implements a servlet for dynamically creating content based on the interaction with the engine invoked by the main page of the SISCO.web system. The SISCO Rules Data Receiver and the SISCO Rule Data Sender components, finally, act as interfaces with the message broker, transforming the asynchronous communication with the broker into the classic synchronous request/response client/web server communication.

---

[10] This guideline has been updated during the publication of the sixth edition of ICD-10 in 2019 and later with the publication of ICD-11 release.

[11] More details on the Drools key components can be found at Red Hat, Inc., Drools rule engine. Full documentation section: <https://www.drools.org/>.

The decision tree represents knowledge in the form of "if P then Q" rules. In the decision tree diagram, non-leaf nodes have two outgoing arcs: YES and NO. The rules defined for each node determine the selection of the outgoing arc and, consequently, the next computed node, based on terminological codes and user responses to the engine. The rules defined on two arcs from the same node are mutually exclusive to ensure the path's clarity. The decision algorithm takes two ICD-9-CM code lists as input: Pathological conditions (PC) and Procedures and Interventions (PI). The selection of the outgoing arc can be determined in two ways: automatically, based on the KB terminology codes feeding the engine, or decided by the user if no knowledge is available in the KB. If the rule engine is unable to ascertain the fulfilment of a rule based on incoming terminological codes, or when a decision necessitates the clinician judgment (e.g., *Are the pathological conditions related to each other?*), the engine will prompt user intervention by formulating a question within the web interface. This question may seek a binary YES/NO response (e.g., *Has it caused complications?*) or the selection of one or more terminological codes (e.g., *Identify the most complex event*). Subsequently, the engine will generate a JSON message encompassing all requisite details for presenting the question, including the query text, answer type (binary or selection of codes), and permissible response values (e.g., YES/NO, TRUE/FALSE, or specific codes).

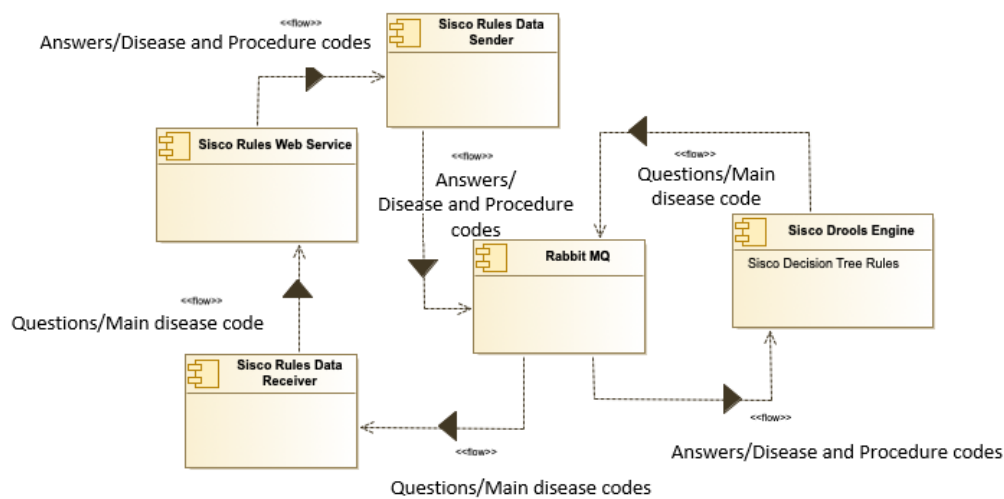

**Figure 4.** The Rules Engine Component Diagram

In this way, The Web User Interface (WUI) content is automatically created by the browser, generating fields based on the answer type. For example, radio buttons are used for exclusive choices and check buttons for multiple choices. Fig. 4 shows two Drools rules for states 18 and 19 in the decision tree diagram. The "S18_ask" rule prompts the user to indicate one or more pathologies not related to the intervention. The "S19_true" rule manages the arrival of the response and determines the next transition from state 19 ("is it a single pathological condition?") based on whether the user has selected one or more codes among the relevant conditions. The result of the rule execution is reaching a leaf node associated with one or more codes suggested for the main pathological condition, which is then displayed in the SISCO.web interface.

```
rule S18_ask
  salience 120
  when
    c:BinaryParameters(i:id, s:state)   from entry-point "entry"
    p:Parameters(ip:id,dd:diagnosis) from entry-point "entry";
    eval(s==18);
    eval(i==ip);
  then
    System.out.println(i +" S18 Identificare una o più condizioni patologiche non correlate all'intervento ");
    JSONObject json = new JSONObject();
    json.put("id", i);
    json.put("state", s);
    json.put("message", "Identificare una o più condizioni patologiche non correlate all'intervento ");
    json.put("type", "ask_multicode");
    CEventsNotification.notify("out.siscoweb.notify_message",json.toString());
    modify(c){state=19}
end

rule S19_false
  when
    c:BinaryParameters(i:id, s:state)   from entry-point "entry"
    ba:StateParameter(sb:state, t:type, cds:codes, ib:id) from entry-point "entry";
    eval(s==19);
    eval(t=="ask_multicode");
    eval(s==sb);
    eval(i==ib);
    eval(cds.size>1);
  then
    modify(c){state=20}
end
```

**Figure 5.** Drools S18-S19 rules example

## 3. Evaluation

After an internal test conducted by the project's informaticians and terminologists, a more detailed usability test was performed by three physicians: This evaluation employed a subset of pathological conditions extracted from the SEI database mentioned in Section 2, along with diseases and interventions drawn from several anonymized patient discharge letters. These LDO contained multiple diagnosis and interventions/procedures, particularly focusing on complex cases characterized by comorbidities and intricate diagnostic definitions. The aim was to assess the tool's effectiveness in suggesting appropriate codes, required for completing the SDO. At this stage, the evaluation was more qualitative than quantitative, as the physicians were unable to access an LDO/SDO database for the project. Nonetheless, initial results indicate that the system performed well, successfully suggesting the most appropriate ICD-9-CM diagnosis even in instances where the input text in the search box of the TEM module was complex or included comorbidities. On average, SISCO.web provided precise ICD-9-CM code suggestions for 80% of 30 use cases tested by physicians, with improved accuracy when using the related terms feature. An example of diagnosis coding (in this case "diabetes") is given in Fig. 6. Here, when a user types "diabete" (diabetes) into the search box, the system auto-completes with suggestions like "diabete-nanismo-obesità" (diabetes-nanism-obesity) and "pre-diabete" (prediabetes). After selecting "diabete", the system displays matching classes in the search results section (considering all the attributes associated to the class, such as title, other description, inclusions, exclusions, alphabetic index terms, etc.), ordered by score. It also suggests related terms (on the left of the page) that co-occur with "diabete" in the ICD-9-CM descriptors. The user can then select a related term like "mellito" (mellitus), prompting the system to refine results based on both selected terms.

At each iteration, the system displays matching classes and related co-occurring terms based on user input. The search progressively narrows down until the user identifies and selects the correct ICD-9-CM class, which is then added to the "Selected Diagnoses" section at the bottom left of the page. Before selecting the proper code, for each code in the results list, the user can view code details (displayed if present on the right of the page and represented by symbols), including:

- *Leaf nodes*: Indicates to select a leaf code from the list presented, being the selected code not a leaf code.
- *Exclusion criteria*: Lists conditions excluded by that ICD-9-CM class.
- *Basic diseases* attribute: Advises coding a basic disease before using the selected code.
- *Use additional codes*: Recommends additional codes relevant to the selected class.

These features resulted helpful for avoiding inconsistencies, providing alerts to key ICD-9-CM coding rules, such as the necessity of coding a basic disease first or using leaf codes instead of general three-digit diagnosis codes (rules which in many cases are unknown by professionals or in some cases taken for granted).

**Figure 6**. SISCO.web Interface: An example of coding for "diabetes mellitus" diagnosis.

Not completely known is also the need for the combined use of the alphabetical and the systematic index of ICD-9-CM (both part of the KB) which help in extending knowledge about a code, providing references to additional codes related to the selected one, etc. Another useful feature of the TEM module was considered the possibility to show, starting from an ICD-9-CM class in the search results, the hierarchy of the classes, derived from the ICD-9-CM ontology, including all the details for each code.

Regarding the second module focused on the identification of the main condition (IMCM) the WUI, illustrated in Fig. 7, consists of three main sections: the upper section displays the two lists of codes (for diagnoses and procedures) selected by the user in the TEM module; the central section interacts with the user during the decision tree process, and the lower section reveals the main diagnosis once it has been identified.

**Figure 7.** SISCO.web Interface: Rule engine support to identify the main condition.

When the user opens the module page, he will see two lists of codes at the top and a progress bar further down. At this point, the backend navigates the decision tree until it hits the first node that requires user input. At this stage, the rule engine requests the user input the necessary parameters to continue the navigation of the tree. These may include, for instance, the "most resource-consuming pathological condition during hospitalization" among the coded diagnoses (in case of multiple diagnoses). The user then selects one from a combo box, thereby entering the required parameter into the module. Subsequently, the rule engine resumes the path of the tree until the final node is reached, i.e., the identification of the main condition, which is finally displayed to the user for confirmation via a dedicated button.

The central part of the page displays only a partial representation of the decision tree structure, including nodes requiring manual input, and the final three stages. This should help the user understand the operations performed by the rule engine to determine the main diagnosis. Nevertheless, the system can autonomously perform certain steps in the decision tree, utilizing previously provided information, the formalized coding rules, and inferences derived from the KB.

Furthermore, the WUI provides a button that cancels the rule engine operations and returns to the text encoding module WUI.

The SISCO.web system was tested both in terms of the usability and efficiency of the search algorithms, by the doctors involved in the project, and in terms of functionality and performance, by the team of computer experts and terminologists who developed the service. The test highlighted that the search results for ICD-9-CM diagnoses obtained using the mentioned algorithms are substantially superimposable. However, it is noted that:
- The weights assigned to the various ICD-9-CM attributes associated with each ICD class in the search results appear inconsistent concerning the relationship between the importance of the various resources present in the KB and the recurrence of the terms (roots);
- Hierarchical algorithm guarantees greater appropriateness in the selection of ICD-9-CM categories since it maintains the relationship of importance between the resources present in the KB even in the event of their enrichment.

Hence, it was necessary to refine the weights assigned to the various attributes[12] to guarantee appropriateness in the selection of ICD-9-CM categories even in the event of moving KB resources

---

[12] In particular, weights ranges from 0 to 10: the main description of the ICD class in the systematic index was still considered the most important with weight 10, the additional terms of the ICD class title have weight 7,5; inclusion terms have weight 2,5; alphabetical

from one step to another. The steps of the algorithm implemented for the coding activity of a diagnosis were confirmed.

The test revealed issues in the identification of the diagnosis module, which is almost related to the formalization and computerization of the decision tree, particularly for some steps of the tree where the physician's input is necessary. This is especially true when the physician selects multiple interventions, as it's crucial, at a certain point of the process to indicate the relevance of each one. The decision tree is not fully computerized in terms of additional resources for automating certain steps (as it can be for example a list of relevant interventions aligned to anatomical sites or mapped to diagnosis categories, which although available in pdf, is still under elaboration for the integration into the rule engine) and allowing the physician to select multiple options. Currently, the computerized decision tree enables the physician to identify the main pathological condition by answering a series of YES/NO questions.

## 4. Related works

Different coding support systems have been developed in the last two decades. Some of them aimed to support the coding of causes of death, generally coded using ICD-10. Examples of these tools are MICAR-ACME, of the US National Center for Health Statistics (Israel 1990), and the IRIS system developed by a European consortium (Pavillon *et al.* 2005). The main issue encountered in these systems is the processing of natural language, which, in the last twenty years has been faced with developing automated coding tools based on Natural Language Processing (NLP) algorithms (see Friedman *et al*. 2004). Only a few systems were based on properly defined coding rules, as done by (Farkas and Szarvas 2008) and (Cardillo *et al.* 2018), both focused on the ICD-9-CM coding. In recent years, challenges have been encountered, from the perspective of Artificial Intelligence (AI) and NLP, based on the literature. Many researchers and companies started applying more sophisticated methods such as Neural Networks or Large Language Models (LLM) to enable Electronic Medical Record (EMR) data coding (Rios and Kavuluru 2018). This trend is confirmed also by the results of the CLEF ICD10 task[13], held in 2020, focused on ICD-10 coding for clinical textual data in Spanish and including, in particular, two subtasks for evaluating systems that predict ICD-10-CM (diagnostic) and ICD-10-PCS (procedural) codes using the CodiEsp corpus (a collection of 1,000 clinical case reports written in Spanish). Here most of the participants used Machine learning approaches and deep learning language models (preferring fine-tuned Multilingual BERT), but also not Machine Learning approaches such as Named Entity Recognition (NER), but the highest mean average precision (MAP) for the prediction of ICD-10 diagnostic codes (0.593) resulted by the combination of a XGBoost classifier and a Jaro Winkler string matching system (see Miranda-Escalada *et al.* 2020). Other studies focused on the application of general-purpose LLMs (such as ChatGPT 3.5 or 4, LLAMA, etc.) to test their performances in the task of automated coding of diagnoses extracted from Discharge summaries by using the ICD-10 classification. Nevertheless, gaps between the current deep learning-based approach applied to clinical coding and the need for explainability and consistency in real-world practice were reported (Dong *et al*. 2022). Some studies indicate alternative methods or frameworks specifically designed for automatic ICD coding. For example (Chao-Wei Huang 2022) used a pre-trained language model (PLM) for ICD coding, sharing a similar idea with BERT-XML, an extension of BERT designed for ICD coding. This model was pre-trained on a large collection of EHR clinical notes using an EHR-specific vocabulary (Zhang *et al.* 2020). Additionally, (Kim and Ganapathi 2021) introduced the Read, Attend, and Code (RAC)

---

index "entry term" has weight 2,5, while its indentations (from the first to the sixth one) were assigned weight 0,1; neoplasm entry term in the alphabetic were assigned weight 2,5, and indentations had 0,1; the main description of diagnoses /procedures derived from the other glossaries in the KB, such as MeSH terms, Emergency glossary, General Practitioner glossary, Rare disease terms, etc., were assigned 0,1, and finally all the other attributes had 0.
[13] CLEF eHealth 2020 – Task 1: Multilingual Information Extraction, available at: https://clefehealth.imag.fr/clefehealth.imag.fr/index135c.html?page_id=187

framework for accurate ICD code prediction. Another approach involved the use of off-the-shelf pre-trained generative LLMs to perform ICD coding, without labelled training examples and leveraging the hierarchical nature of the ICD ontology, thus relying on dynamic searches for clinical entities within the ICD ontology (Boyle *et al.* 2023).

One issue that needs to be highlighted in this context is the lack of available datasets for ICD coding to train AI-based models, especially in some languages, such as Italian. Few approaches in the literature show how to mitigate this problem. For example, (Almagro *et al.* 2019) propose a cross-lingual approach based on Machine Translation methods to code death certificates with ICD-10 through supervised learning. This means that they tried to code Italian death certificates using certificates from another language (French), so combining collections of different languages to increase the availability of coded documents. Improvements in the system performance here were observed for codes assigned to labels with few occurrences. Silvestri et al. (2020) conducted a study on cross-lingual XLM fine-tuning aimed at predicting and classifying ICD-10 codes. They provided a preliminary evaluation of a model fine-tuned on short medical notes written in English using an Italian language test set. Even though, the results indicated the need for further experiments to increase the number of samples in the test set (using various document collections), to better assess the model's ability to generalize.

A more recent overview on the topic is provided by the study conducted by the Icahn School of Medicine at Mount Sinai in New York revealed significant shortcomings in the performance of LLMs in clinical coding. The analysis showed that the existing models, including the highest-performing GPT-4, achieved less than 50% accuracy in matching medical codes to clinical texts. Such inaccuracies can result in serious billing errors and compliance issues within healthcare systems. Furthermore, the study highlighted varying performance levels among different LLMs, posing challenges in clinical environments where precise coding is essential for billing and ensuring accurate patient care. (Soroush *et al*. 2024).

These results emphasize the necessity for refinement and validation of these technologies before considering clinical implementation, thus providing customized AI tools specifically designed for medical coding, instead of using general-purpose LLMs.

Given this overview, we can state that SISCO.web performances are comparable with most of the mentioned systems. Unlike existing systems and the most recent AI-based coding support, SISCO.web offers dual support. Firstly, it helps in finding the appropriate ICD-9-CM (or in the future ICD-10) code for a diagnosis or procedure by utilizing NLP techniques combined with the application of trustworthy coding rules, which are necessary to know when dealing with the selected classification system. Secondly, it assists in identifying the main diagnosis (the most serious and/or resource-intensive during hospitalization or the inpatient encounter) among multiple diagnoses, which is often a challenging and underestimated task. The advantages of this approach also stem from the integration of decision tree algorithms, which expand the system's functionalities.

## 5. Conclusions and future directions

This paper shows the approach used for the development of a web service that provides support to physicians, particularly hospital doctors, during the compilation of the SDO while coding in ICD-9-CM, and where necessary in ICD-10, for the main pathology, secondary pathologies, and procedures and interventions. The system also proposes a module based on a series of formal rules that represent a decision tree specifically designed for identifying the main pathological condition, which needs to be indicated and coded in a separate field in SDO. The evaluation of the text encoding module, allowing for the search and suggestion of ICD-9-CM coding for diseases and procedures, has reached fairly good performances in terms of the accuracy of the coding suggestions, the efficiency of the system, and also regarding the usability of the system in the reference context. Differently, some limitations are highlighted concerning the rule engine module, which allows, through a series of steps and interactions with the user, the identification of the main diagnosis. In this case, the initial formalization of the rules provided by the decision tree did not yield the expected results. It has therefore become necessary to update the rules and, above all, to make available ad hoc terminological resources to be submitted to the rule engine to automate certain steps of the decision tree, thus ensuring the required performance compared to other support systems available in the literature. Considering that ICD-9-CM is currently mandatory in Italy for coding diagnosis into hospital discharge records, the prototype and tests of SISCO.web uses this ICD version to be used in hospital coding. Nevertheless, the system has been designed to work using different ICD versions, thus also ICD-10, including the design of a decision tree for identifying the main diagnosis set specifically for ICD-10. This possibility, recently, resulted advantageously since, as mentioned in Section 1, the Italian Ministry of Health, to be aligned to European guidelines on cross-boarding care, is working on a roadmap to shift from ICD-9-CM to ICD-10-IM for the coding of morbidities in SDO, leveraging the results of the It.DRG project. For this reason, future work will be the extension of the system, in terms of integration of the KB with the Italian version of ICD-10 (the mentioned ICD-10-IM) and the necessary crosswalking tables as well as the implementation of the already defined ICD-10-based decision tree in the rule engine. At the same time, it will be possible to set up versions of this support system able to manage classifications of interventions other than those used in Italy. Another possible future work is the development of a JavaScript library to distribute the service to interested parties and test it on a large scale (i.e., some hospital wards). As observed in Section 4, automated clinical coding holds promise for AI despite the technical and organizational challenges, but coders need to be involved in the development process, as done in the present work. Given this understanding, it can be argued that SISCO.web could serve as a good compromise, particularly if focusing on a new research direction that could be pursued over the next five years. This would involve improving the approach using LLMs + Retrieval-Augmented Generation (RAG) to enhance both the text encoding module and the implementation of the decision tree in the rule engine. Another possible future work could be to use a complementary approach for the analysis, through NLP/DL, of the diagnostic sections of hospital discharge letters (LDOs in Italy), which are very detailed reports. In our use case, a sample of these documents was used to test the performances of SISCO.web in terms of capacity to support coding for complex search records (e.g., comorbidities, very detailed diagnoses, etc.). In the future, it would be valuable to explore the possibility of providing coding support while registering LDOs' data, particularly in the diagnostic section. This information is crucial for the transition of care from the hospital to the community setting. Additionally, it will be important to optimize the communication between the coding support system and the company's EHR information systems used by physicians.

Titolo